# SUBSTRUCTURES OF CLUSTERS AND COSMOLOGICAL MODELS [*]

Y.P. JING[†], H.J. MO. AND G.BÖRNER
*MPI für Astrophysik, Karl-Schwarzschild-Str. 1, 85748 Garching, Germany*

and

L.Z. FANG
*Department of Physics, University of Arizona, Tucson, AZ 85721, USA*

## ABSTRACT

We investigate, using high resolution N-body simulations, the density profiles and the morphologies of galaxy clusters in three models of structure formation. We show that these properties are closely related to the occurrence of a significant merging event to a cluster in the recent past. The three models are: (1) the standard CDM model (SCDM) with $\Omega_0 = 1$, $\Lambda_0 = 0$ and $h = 0.5$; (2) a low-density flat model (LCDM) with $\Omega_0 = 0.3$, $\Lambda_0 = 0.7$ and $h = 0.75$; (3) a CDM dominated open model (OCDM) with $\Omega_0 = 0.1$, $\Lambda_0 = 0$. We find that the clusters in the OCDM model have the steepest density profiles, their density contours are the roundest and show the smallest center shifts, among the three models. The clusters in the SCDM model have contours that are most elongated and show the largest center shifts, indicating the existence of frequent mergers in the recent past. The clusters in the LCDM model have density profiles and center shifts (in their density contours) similar to those in the SCDM model, although the density contours tend to be rounder. Our results show that, although the density profiles and morphologies of clusters depend on models of structure formation, the LCDM model, which is currently considered as a successful alternative to the SCDM model, can do as well in producing a substantial fraction of clusters with substructures. This is in contrast to the conception that this model may have serious problem in this aspect.

## 1. Introduction

Observational studies have demonstrated that a considerable fraction of clusters show evidence of substructures in the galaxy distribution and in the x-ray images. The observational data are expected to be improved greatly in the near future. These observations are potentially useful in constraining models of structure formation.

According to current models of structure formation, clusters of galaxies, which are the largest collapsed objects in the universe, are expected to be dynamically young, to show substructures and to have density profiles that differ from those expected from dynamical equilibrium. The deviation from a completely relaxed configuration should, however, be different for different models of structure formation, for the characteristic time of cluster formation depends both on the cosmological model and on the power spectrum of the initial density fluctuations. Here we report some results of our recent work[1] (JMBF) on the substructures and density profiles of clusters in three models of galaxy formation.

## 2. Cosmological models and N-body simulations

We have run simulations for the Standard CDM model (SCDM) and for a low-density flat CDM model with a non-zero cosmological constant (LCDM). The SCDM model is assumed to have $h = 0.5$,[‡] $\Omega_0 = 1$ and $\Lambda_0 = 0$; the LCDM model has $h = 0.75$, $\Omega_0 = 0.3$ and $\Lambda_0 = 0.7$. For both models we assume a primordial power spectrum with the Harrison-Zel'dovich form. The initial spectra are normalized, so that the SCDM model has $\sigma_8 = 0.6$ and the LCDM model has $\sigma_8 = 1.0$, where $\sigma_8$ is the *rms* linear density fluctuation in a sphere of radius $8\,h^{-1}$Mpc at $z = 0$. The LCDM model is currently interesting because it is compatible with most observational data. Here we have also run simulations for a CDM dominated open universe (OCDM). In this case, we take $\Omega_0 = 0.1$, $\Lambda_0 = 0$ and $\sigma_8 = 1$. The initial power spectrum used for this model is the same as that for the SCDM model. The parameters for the OCDM model are neither theoretically nor observationally motivated. We take this model as an extreme case to show the dependence of cluster internal structures on cosmological parameters.

The simulations were performed by using a P$^3$M code in a cubic box of size $128\,h^{-1}$Mpc, with force resolution $\eta = 0.1\,h^{-1}$Mpc. We have run 3 realizations for the SCDM model. In the first two runs, $100^3$ particles were used, and in the third run we use $128^3$ particles. For the LCDM and OCDM models, we have run five realizations and use $64^3$ particles.

To identify cluster-like dark halos, we use the same procedure as described in Ref. [2]. Briefly, we first use the *friends-of-friends* algorithm to find groups. The groups identified in this way can have irregular shapes. Some of the groups may have their centers of mass not located in dense regions. Since an accurate determination of the centers of dark halos is important in our analysis, we treat this kind of situation by searching for potential minima around the groups found by the *friends-of-friends* algorithm. The final accuracy of our center determination is better than $0.05\,h^{-1}$Mpc.

## 3. The density profile and morphology of clusters

Here we use four different measures to quantify the density profiles and morphologies of clusters.

*3.1. The density profiles of clusters*

To quantify the density profiles of clusters, we measure the cross-correlation function $\xi_{cm}(r)$ between cluster and mass particles. We find that the density profiles are much steeper in the OCDM than in the SCDM and LCDM models. The latter two

---

[‡]The Hubble constant $H_0$ is written as $H_0 = 100\,h\,\mathrm{km\,s^{-1} Mpc^{-1}}$

models have nearly the same shapes in their density profiles. The cross-correlation functions can be well described by power-laws $\xi_{cm}(r) \propto r^{-\gamma}$ for $r$ less than a few $h^{-1}$Mpc. By fitting $\xi_{cm}(r)$ in the range between $r = 0.2$ and $1.0\, h^{-1}$Mpc, we find that the slopes $\gamma$ are $2.34 \pm 0.05$, $2.30 \pm 0.05$ and $2.9 \pm 0.1$ for the SCDM, LCDM and OCDM models, respectively. The $\gamma$ values depend weakly on mass or velocity dispersion of the clusters.

*3.2. The shape of virialized halos*

Assuming that the isodensity surfaces of a cluster are approximately described by a triaxial ellipsoid, we calculate the axial ratios $a_1/a_3$ and $a_2/a_3$ ($a_1 \leq a_2 \leq a_3$). Our calculation is done for the virialized part of a cluster. The axial-rato distributions of the three models have similar widths. The mean values of $(a_1/a_3, a_2/a_3)$ are (0.46, 0.63) for SCDM, (0.47, 0.65) for LCDM, and (0.58, 0.73) for OCDM. The clusters in the OCDM model have systematically higher axial ratios than in the other two models, which means these clusters are systematically rounder. The axial-ratio distributions for the SCDM and LCDM models are quite similar.

*3.3. Cluster shapes in projection*

We study further the morphology of clusters from their projected density distribution. We first obtain a smoothed 3-D density distribution in a cube of side $3.2\, h^{-1}$Mpc with a gaussian smoothing kernel. We then calculate the square of the density in each cell and project the density square on a given plane. The reason for using density square in the projection is to have a more direct comparison with the x-ray surface density (the x-ray luminosity in a cell is proportional to the square of the gas density in the cell). Very similar results were obtained when we used density itself in the projection. For brevity, we will denote the projected field of density square by $\mathcal{D}S$.

Figure 1 shows the contours of the $\mathcal{D}S$ distributions for the four most massive clusters of in one realization of the three models. As shown in JMBF, the $\mathcal{D}S$ contours clearly reflect recent merging events. Inspecting particle distributions, we find that each of the three clusters, cl1, cl3 in LCDM and cl4 in SCDM, has a significant merging event in the recent past ($z \lesssim 0.3$). Their $\mathcal{D}S$ contours all show peanut-like shapes near their centers and have significant center shifts. In contrast, clusters like those in OCDM, which show round contours with small center shifts, have no significant recent mergers. The distortions in the $\mathcal{D}S$ contours, visible for cl1 and cl2 in the SCDM model and for cl2 in the LCDM model, are due to recent small mergers. Therefore, the shape and the center shift of the $\mathcal{D}S$ contours can be used as an indicator of the dynamical state of clusters.

In Figure 2 we show the distributions of the axis ratios and center shifts for the 50 clusters of the highest velocity dispersions in each realization of the three models. From the figure, it is clearly seen that most clusters in the OCDM model are very round, with axis ratios bigger than 0.8. The center shifts in this model

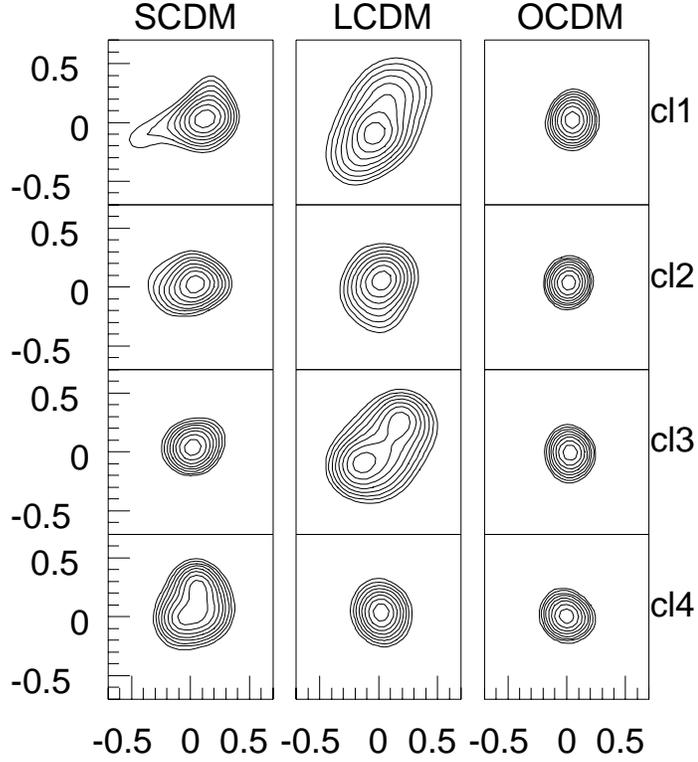

Fig. 1. The projected density-square contours of twelve clusters. The $i$-th contour level is chosen to be $10^{-0.05\times(3i-2)}$ times the maximum $\mathcal{DS}$ value. Scale units are in $h^{-1}\mathrm{Mpc}$

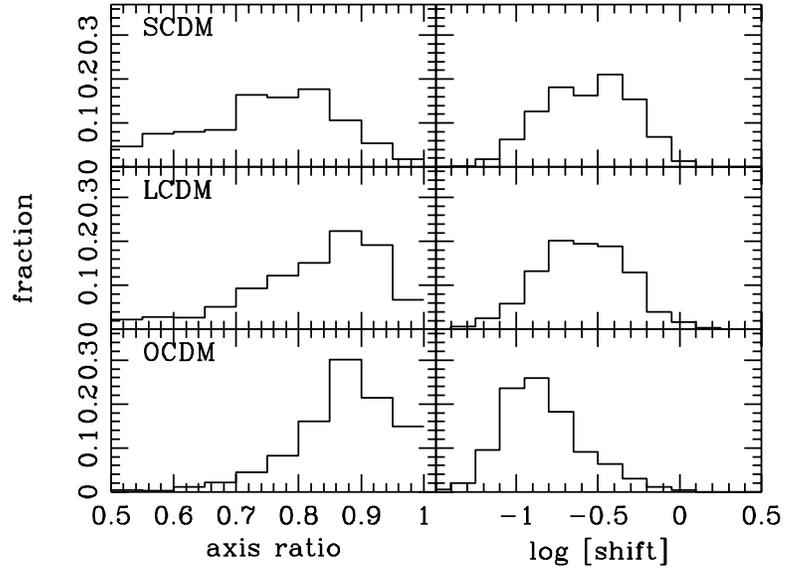

Fig. 2. The distributions of the mean axis ratios and center shifts of the projected isodensity contours in a central circular region with radius $0.5\,h^{-1}\mathrm{Mpc}$ around clusters. Shift unit is in $50\,h^{-1}\mathrm{Kpc}$

are also significantly smaller than those in the other two models. The center shift distributions for both SCDM and LCDM models are quite similar, while clusters in the LCDM model appear rounder, as shown by the axis ratio distributions.

*3.4. X-ray images using hydrostatic models*

We construct x-ray images for a cluster under the following assumptions: 1) the intracluster gas is in hydrostatic equilibrium under the gravitational potential of dark matter; 2) the gas is isothermal; 3) the central gas temperature is equal to the virial temperature; and 4) the gas composition is primordial. Inspecting the x-ray iso-intensity contours, we find that all important features seen in the $\mathcal{DS}$ contours (Fig. 1) are also seen in the x-ray contours, though the x-ray contours are rounder than the $\mathcal{DS}$ contours, as expected. Our quantitative study of the axial ratios and center shifts of the x-ray contours gives a result similar to Fig. 2.

## 4. Discussion

Based on 8 clusters from N-body gasdynamic simulations, Evrard et al.[3] found that the x-ray images of clusters in low-density models (with $\Omega_0 = 0.2$) are much more regular, spherically symmetric and centrally condensed than those in an Einstein-de Sitter model ($\Omega_0 = 1$), with only a weak dependence on a possible cosmological constant. Their result seems in contrast with our results on the LCDM model. We found that the clusters in this model have density profiles and center shifts of contours that are very similar to those in the SCDM model. The density and x-ray image contours are indeed rounder than in the SCDM model. However we do see a large fraction of clusters in the LCDM model showing significant substructures. It is still unclear to us what causes this discrepancy. The apparent differences between the two studies are: (1) They have assumed a high fraction (50 percent) of mass in baryons for their low-density models, which might have the effect of making the x-ray images rounder. (2) Since their simulation boxes are small, the periodic boundary condition used in their code may reduce the clustering power on large scales. This effect might be more severe in the low-density models. (3) The values of $\Omega_0$ are 0.2 in their work and 0.3 in our work. This may account for some of the difference. (4) In our work we have based our analysis on the dark-matter distribution. Some of the discrepancy may be explained, if the x-ray emission does not trace well the potential of dark matter. Some of our assumptions in constructing the x-ray images may not be valid. The polytropic model may erase some small components. However, we expect that the real x-ray images will have properties between the $\mathcal{DS}$ and our x-ray images.